\documentclass[aps,pra,reprint,superscriptaddress,floatfix]{revtex4-2}

\usepackage{xcolor}
\usepackage{amsmath}
\usepackage{amssymb}
\usepackage{graphicx}
\usepackage{hyperref}
\usepackage{xcolor}
\usepackage{bm}
\usepackage{siunitx}
\begin{document}
\title{Magneto-optical study of Nb thin films for superconducting qubits}
\author{Amlan Datta}
\affiliation{Ames National Laboratory, Ames, IA 50011, USA}
\affiliation{Department of Physics \& Astronomy, Iowa State University, Ames, IA
50011, USA}
\author{Kamal R. Joshi}
\affiliation{Ames National Laboratory, Ames, IA 50011, USA}
\author{Sunil Ghinire}
\affiliation{Ames National Laboratory, Ames, IA 50011, USA}
\affiliation{Department of Physics \& Astronomy, Iowa State University, Ames, IA
50011, USA}
\author{Makariy A. Tanatar}
\affiliation{Ames National Laboratory, Ames, IA 50011, USA}
\affiliation{Department of Physics \& Astronomy, Iowa State University, Ames, IA
50011, USA}
\author{Cameron J. Kopas}
\affiliation{Rigetti Computing, Berkeley, CA 94710, USA}
\author{Jayss Marshall}
\affiliation{Rigetti Computing, Berkeley, CA 94710, USA}
\author{Josh Y. Mutus}
\affiliation{Rigetti Computing, Berkeley, CA 94710, USA}
\author{David P. Pappas}
\affiliation{Rigetti Computing, Berkeley, CA 94710, USA}

\author{Matthew J. Kramer}
\affiliation{Ames National Laboratory, Ames, IA 50011, USA}

\author{Ruslan Prozorov}
\email{[corresponding author] prozorov@ameslab.gov}
\affiliation{Ames National Laboratory, Ames, IA 50011, USA}
\affiliation{Department of Physics \& Astronomy, Iowa State University, Ames, IA
50011, USA}

\date{10 February 2026}

\begin{abstract}
Among the recognized sources of decoherence in superconducting qubits,
the spatial inhomogeneity of the superconducting state and the possible
presence of magnetic-flux vortices remain comparatively underexplored.
Niobium is commonly used as a structural material in transmon qubits
that host Josephson junctions, and excess dissipation anywhere in
the transmon can become a bottleneck that limits overall quantum performance.
The metal/substrate interfacial layer may simultaneously host pair-breaking
loss channels (e.g., two-level systems, TLS) and control thermal transport,
thereby affecting dissipation and temperature stability. Here, we use
quantitative magneto-optical imaging of the magnetic-flux distribution
to characterize the homogeneity of the superconducting state and the
critical current density, $j_{c}$, in niobium films fabricated under
different sputtering conditions. The imaging reveals distinct flux-penetration
regimes, ranging from a nearly ideal Bean critical state to strongly
nonuniform thermo-magnetic dendritic avalanches. By fitting the measured
magnetic-induction profiles, we extract $j_{c}$ and correlate it
with film physical properties and with measured qubit internal quality
factors. Our results indicate that the Nb/Si interlayer can be a significant
contributor to decoherence and should be considered an important factor that must be optimized. 
\end{abstract}
\maketitle

\section{Introduction}

Superconducting qubits with Nb film circuitry are among the best
studied candidates for scalable quantum computers \citep{
,Huang2020,Siddiqi2021}. Niobium has the highest superconducting transition temperature ($T_{c}\approx9.4\,\text{K}$) of all elements, and its performance in various superconducting applications is well established. High-coherence three-dimensional (3D) superconducting radio-frequency (SRF) cavities achieve internal
quality factors, $Q_{i}$, corresponding to photon lifetimes exceeding
2 seconds \cite{alex20}. However, devices based on Nb thin films show lower values, and there is an intense research effort to find and mitigate sources of quantum decoherence. Indeed, thin films have more potential ``bottlenecks'' and require a deeper
understanding of how microstructure, substrate, interfaces, surface oxides, and deposition parameters influence dissipation at microwave frequencies \citep{Kjaergaard2020,Devoret2013,Leon2021,Siddiqi2021,ghimire2024quasi,Joshi2023,Zarea2023,oh2024,mcfadden2025}. 

Among established sources of decoherence, such as two-level systems
(TLS) \cite{Bafia2024, zarea2025, Siddiqi2021, faoro2012, Premkumar2021}, the inhomogeneity of the superconducting state is practically unexplored, except for a few recent works \cite{Joshi2023,oh2024, Murthy2025}. It is not easy to study the superconducting state at the mesoscale. Scanning techniques cannot reveal large areas and dynamic properties, and macroscopic measurements average the signals from many contributions. Here, we utilize the dynamics of the vortex lattice to bridge this gap. We must make it clear: of course, transmon qubits are shielded from external magnetic fields and will never experience even moderate levels. We use the vortices as a sort of litmus test to detect possible issues with the superconducting state. Specifically, vortex behavior is determined by two main quantities: pinning, which originates from defects and impurities, and condensation energy, which is proportional to the magnitude of the order parameter (squared). The analysis of the vortex density distribution is a direct probe of the superconducting state's homogeneity. 

In thin films, vortices can also be used to probe the thermal properties of the Nb/substrate interlayer. It is particularly useful in niobium films, which develop thermal instabilities when the heat generated by moving vortices cannot dissipate properly. The latter is sensitive to the atomic structure of the interface between niobium metal and the substrate. Indeed, the Nb-Si interface has recently been identified as a crucial loss channel in Nb transmon qubits \citep{oh2024, lu2022, Mueller2019, woods2019}. Cross-sectional transmission \citep{BlancoAlvarez2019,Tanatar2022,Jing2015,ghimire2024quasi,Joshi2023} electron microscopy and spectroscopy have revealed that Nb deposited on Si forms a several-nanometer-thick intermixed layer consisting of amorphous and crystalline niobium silicides \citep{oh2024, lu2022}. This region can have a range of atomic structures, from disordered to well-crystallized, different stoichiometries, and different thicknesses, all of which depend sensitively on how the film was made. 

Of course, the study of vortex physics must be complemented by other measurements and characterizations as well as measurements of the quality factors in devices made from the Nb films of interest. In this work, we utilize magneto-optical imaging to investigate vortex behavior in Nb thin films fabricated on silicon substrates at various deposition rates and establish a correlation with the measured internal quality factors. This approach provides a rapid and efficient method for delivering prompt feedback to optimize the film's fabrication process.

\section{Samples}

Niobium (Nb) films were deposited on high-resistivity silicon wafers
($\rho\geq10\,\text{k}\Omega\cdot\text{cm}$) using physical vapor
deposition (PVD). 
\begin{itemize}
\item \textbf{Sample A:} fabricated using high-power impulse magnetron sputtering
(HiPIMS). 
\item \textbf{Sample B:} fabricated using DC sputtering with a low-high
power sequence. 
\item \textbf{Sample C:} fabricated using DC sputtering at high power. 
\end{itemize}

All samples underwent a 5:1 buffered oxide etch (BOE) dip and identical pre-treatment, followed by an \textit{in situ} bake prior to Nb deposition.
The sputter targets were 99.999\% pure Nb, and the nominal film thickness was 175 nm. Both HiPIMS and DC sputtering processes used the same
substrate rotation and argon pressure \citep{oh2024, Nersisyan2019}. The HiPIMS process applied 850~V voltage pulses with a 1\% duty cycle, yielding a deposition rate of approximately 5.1~nm/min. In comparison, DC sputtering produced deposition rates of 5.3~nm/min at 75~W (low power) and 25~nm/min at 350~W (high power). For Sample B, approximately 30~nm was deposited at low power before switching to high power to complete the film thickness, chosen to match the HiPIMS deposition rate.

\section{Magneto-optical imaging}

Polarized light optical measurements at low temperatures were performed in a closed-cycle optical cryogenic station (\textit{Montana Instruments}, Model S50) integrated with an \textit{Olympus BX3M} polarized light microscope equipped with long working distance objectives. The microscope-cryostat combination provided stable optical access over a wide temperature range while maintaining excellent mechanical and thermal stability, which is essential for high-resolution imaging at cryogenic temperatures. The sample was rigidly mounted on a gold-plated copper stage inside the vacuum sample chamber. This closed-cycle $^{4}$He refrigeration system enables continuous measurements from room temperature down to 3.5~K.

Real-time spatial mapping of the normal component of the magnetic induction was performed using magneto-optical (MO) imaging based on the Faraday effect. In this technique, a transparent ferrimagnetic indicator is placed directly on the sample surface. The indicator is a multilayer film with an active layer of bismuth-doped ferrimagnetic iron-garnet with in-plane magnetization, a mirror, and a protective layer. The indicator is placed mirror-side down so that the light reflected off the mirror never touches the sample. The linearly polarized light propagates through the garnet and picks up a double Faraday rotation. The intensity of the light component that passes through the linear analyzer is proportional to the local $B_{z}\left(r\right)$, enabling local sensing of the perpendicular component of the magnetic induction, $B_{z}\left(r\right)$, with high sub-$\mu$m spatial and temporal resolution. In the images in this paper, brighter areas correspond to a higher amplitude of $B_{z}\left(r\right)$. A detailed description of this technique can be found elsewhere ~\citep{Young2005,Prozorov2006a,Jooss2002}. More recent applications to study magnetic induction in niobium thin films can be found in Refs.\citep{Joshi2023,Datta2024}.

\section{Results and discussion}

Magneto-optical (MO) imaging was performed at zero-field cooled (ZFC) and field-cooled (FC) conditions. In the former, the sample is cooled to a target temperature (in this paper, 4\,K and 6\,K) below the transition temperature, $T_{c}$. After the target temperature is reached, an external magnetic field is applied perpendicular to the film's surface, and images are recorded. In the field-cooled protocol, an external magnetic field is applied above $T_{c}$, and the sample is cooled to the desired temperature below $T_{c}$. Once it reaches that temperature, the field is switched off, and we image the trapped flux inside the sample. 

\begin{figure}[tbh]
\includegraphics[width=1\linewidth]{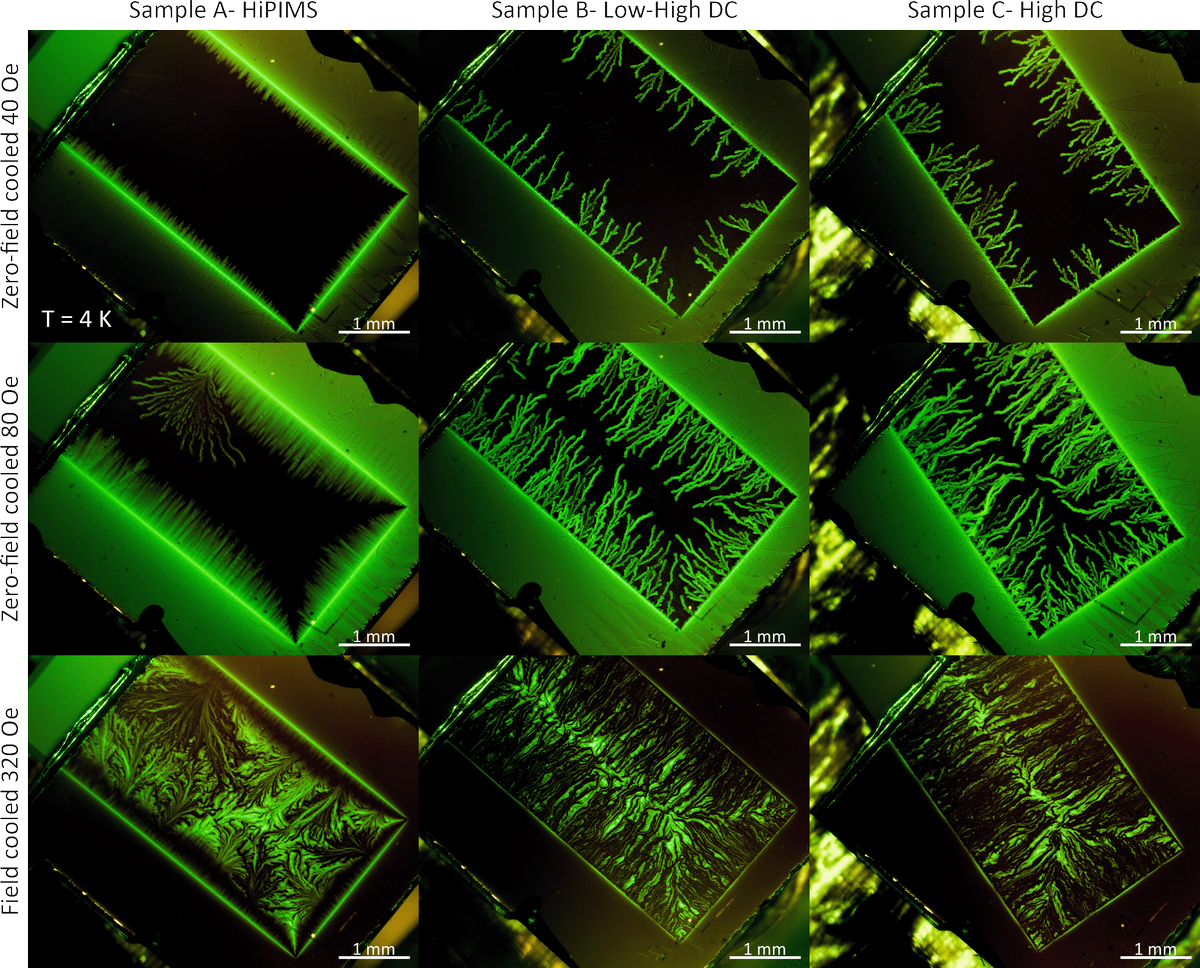} \caption{(Color online) Magneto-optical Faraday images of Nb thin films showing
magnetic flux penetration in sample A-C. Color intensity is proportional
to the magnetic induction on the film surface. The upper two rows are
magnetic field applied in zero-field cooled state at 4\,K and the bottom-most row is showing field cooled images with 320~Oe of applied field at 4\,K. Sample B and Sample C exhibit the presence of thermo-magnetic dendritic avalanches even at an applied field of as low as 40~Oe.}
\label{fig1:MO_4K}
\end{figure}

Figure \ref{fig1:MO_4K} shows ZFC and FC images of three Nb thin film samples deposited under three different conditions. The upper two rows of Fig.\ref{fig1:MO_4K} show flux penetration into the sample when magnetic fields of 40\,Oe and 80\,Oe were applied. Sample A behaves differently from sputtered Samples B and C. Flux penetration proceeds through two distinct flux-dynamic regimes. The first one is led by rapid dendritic avalanches, visually resembling lightning-like branching patterns. These events arise from a thermal runaway process in which vortex motion generates dissipation, locally suppressing the critical current and enabling abrupt flux propagation along narrow channels \citep{Colauto2008,Motta2013,BaruchEl2014,Jing2015,BaruchEL2016,BaruchEl2018,BlancoAlvarez2019,Vestgaarden2018,Colauto2020}. Microstructural inhomogeneities, such as grain boundaries, average grain sizes guide branching and determine the morphology of the avalanche fronts. In sufficiently thick samples, the instability may grow globally, resulting in a collapse of the established Bean critical state \citep{Prozorov2006a}.In the other flux regime, magnetic flux enters more gradually, forming a quasi-static Bean-like profile that reflects the granular structure of the film. The final trapped-flux configuration consists of a mixed uniform and nonuniform pattern, encoding the dynamical instabilities experienced during the field cycle. 

\begin{figure}[tbh]
\includegraphics[width=1\linewidth]{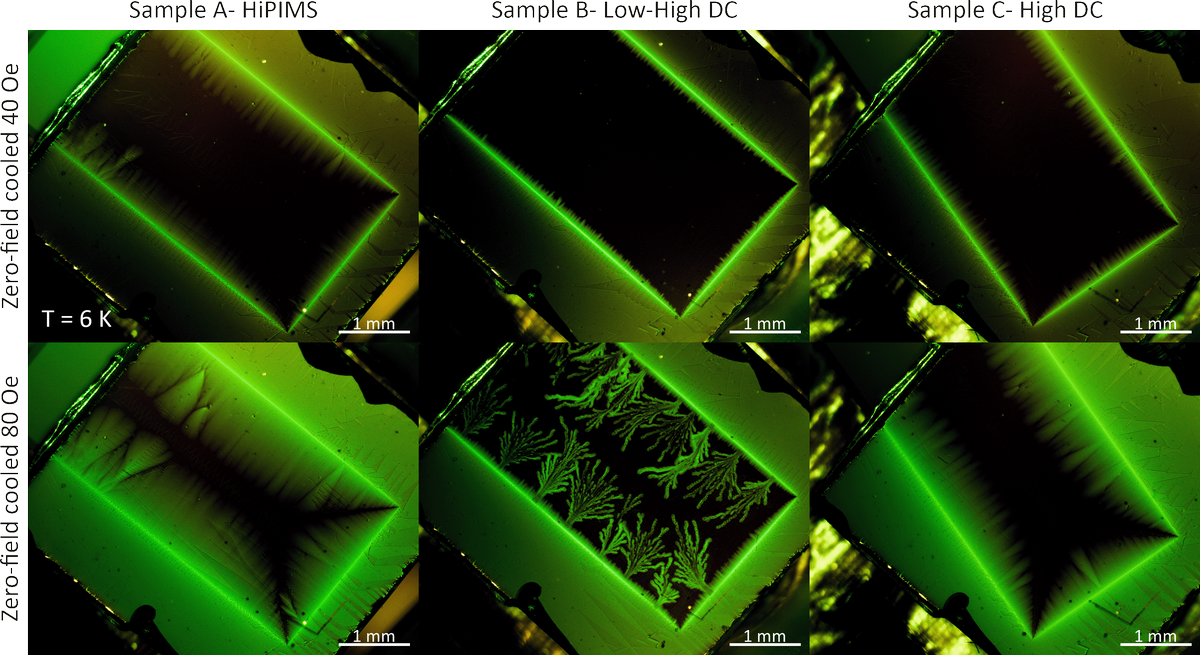} \caption{(Color online) Magneto-optical Faraday images of Nb thin films showing
magnetic flux penetration in sample A-C. Color intensity is proportional
to the magnetic induction on the film surface. Upper two rows are
magnetic field applied in zero-field cooled state at 6\,K. Sample
B still exhibits the presence thermo-magnetic dendritic avalanches even
in zero-field cooled state. This tells about higher flux pinning but
poor thermal coupling in sample B as compared to other films as discussed
in text.}
\label{fig2:MO_6K}
\end{figure}

Dendritic avalanches appear in a limited area of the $T-H$ phase
space, whose size depends sensitively on thermal and magnetic properties
of a studied film. By imaging at a higher temperature, we can tune
to a regime when only some films exhibit dendrites, which will allow
us to distinguish between films. Figure \ref{fig2:MO_6K} shows magnetic flux penetration at $T=6\:\mathrm{K}$. Even though we do not observe
any anomaly with lower magnetic field applied of 40\,Oe, increasing
magnetic field to 80\,Oe triggers the dendritic flux penetration in
sample B. Therefore, we observe distinctly different behavior of three
films. Sample A shows no dendritic avalanches at either temperature.
Sample B displays avalanches at both 4 K and 6\,K, whereas sample
C shows avalanches only at 4\,K. Since the avalanching behavior is
determined by two factors - thermal channeling to a substrate and
the critical current density, $j_{c}$, we need to estimate the latter
and compare the films.

\begin{figure*}[tbh]
\includegraphics[width=1\linewidth]{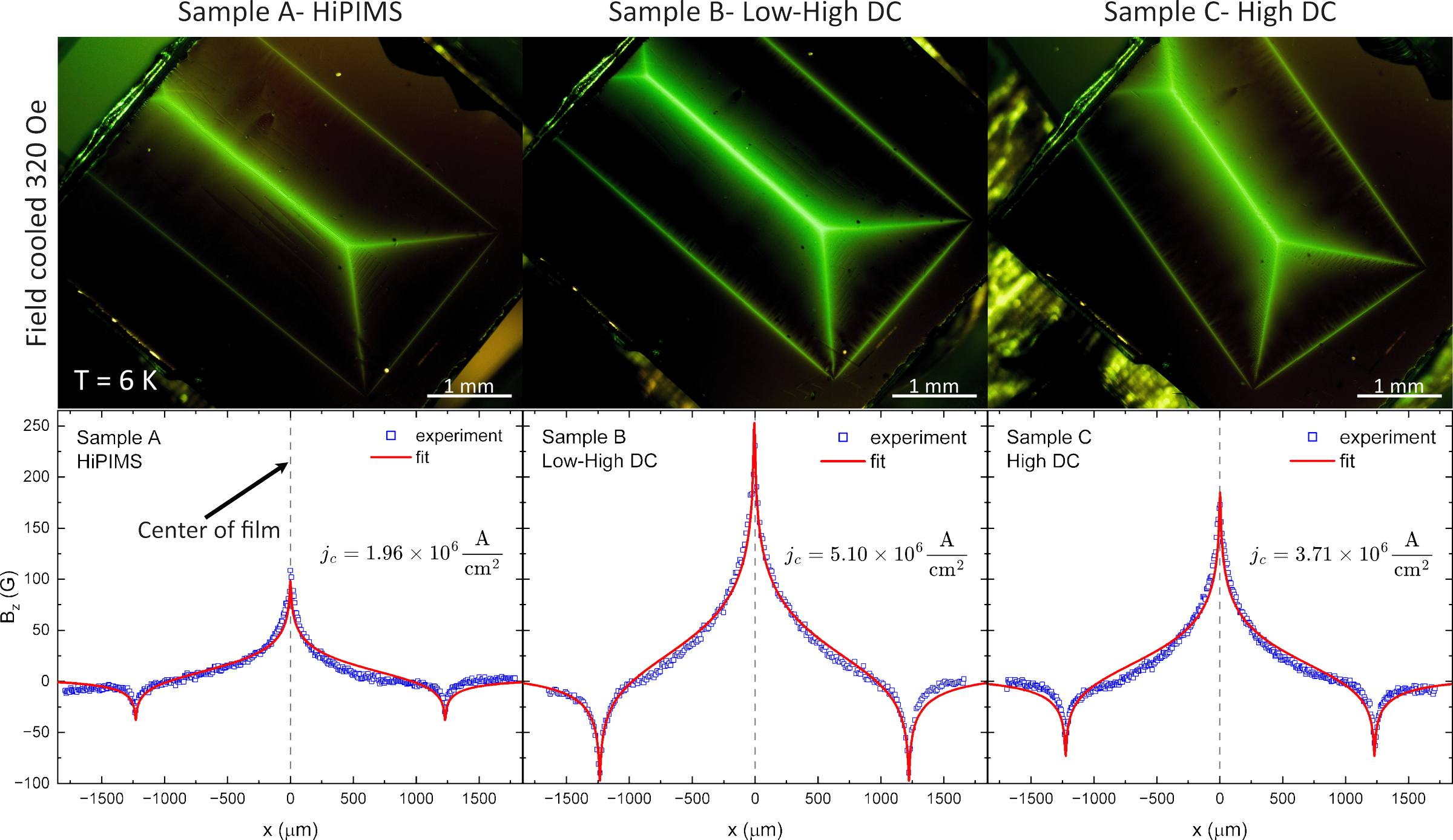} \caption{Magneto-optical Faraday images of Nb thin films showing trapped flux
following Bean's model of flux trapping in field-cooled condition when 320 Oe of magnetic field was applied during the process. The films were field cooled to the temperature of 6\,K as at higher temperatures the films are free of dendritic avalanches and exhibit trapped flux with sharp contrast. Bottom row contains fitting of Eq. with magnetic
induction, calculated from line intensity profile across the film and later converted to magnetic induction, for the estimation of critical current density, $j_{c}$.}
\label{fig3:jc}
\end{figure*}

This can be achieved by analyzing the magnetic induction profiles
in the direction of flux penetration, but the film has to be in a
Bean critical state, and the comparison must be made at the same magnetic
field. It is convenient to examine the remanent state profiles obtained
after cooling in a certain magnetic field to a target temperature
and turning the field off. To obtain the magnetic induction profiles,
an intensity line profile is measured across the film and then converted
to a magnetic induction via the calibration factors determined from
MO images measuring the intensity far from the film at a known applied
field. The top row in Fig.\ref{fig3:jc} shows nearly perfect Bean
roof-like distributions of $B_{z}\left(r\right)$. The elongated rectangular
planar shape of the samples is perfect for an important simplification.
Measuring the profile in the middle of the long side allows us to use the analytical expression for an infinite thin film, 

\begin{equation}
B_{z}\left(x,z\right)=10\,t\,j_{c}\ln\left[\frac{\left((d+x)^{2}+z^{2}\right)\left((d-x)^{2}+z^{2}\right)}{(x^{2}+z^{2})^{2}}\right]\label{eq:jc_fit}
\end{equation}
where the magnetic induction, $B_{z}\left(x,z\right)$, is in gauss,
thickness $t$ is in $\mu\mathrm{m}$, and critical current density,
$j_{c}$, is in units of $10^{6}\,\mathrm{A/cm^{2}}$. The film width
is $2d$, $x$ is the coordinate along the line profile, and $z\thickapprox5\;\mu\mathrm{m}$
is the estimated location of our magneto-optical indicator above the
film surface. The fitted red curves show good agreement with the theory, supporting the validity of our assumptions. The lower panel of Fig.\ref{fig3:jc}
shows the corresponding fitting of Eq.\ref{eq:jc_fit} with $j_{c}$
as the only free parameter. Its values are shown on the same frames. 

\begin{figure}[tbh]
\includegraphics[width=1\linewidth]{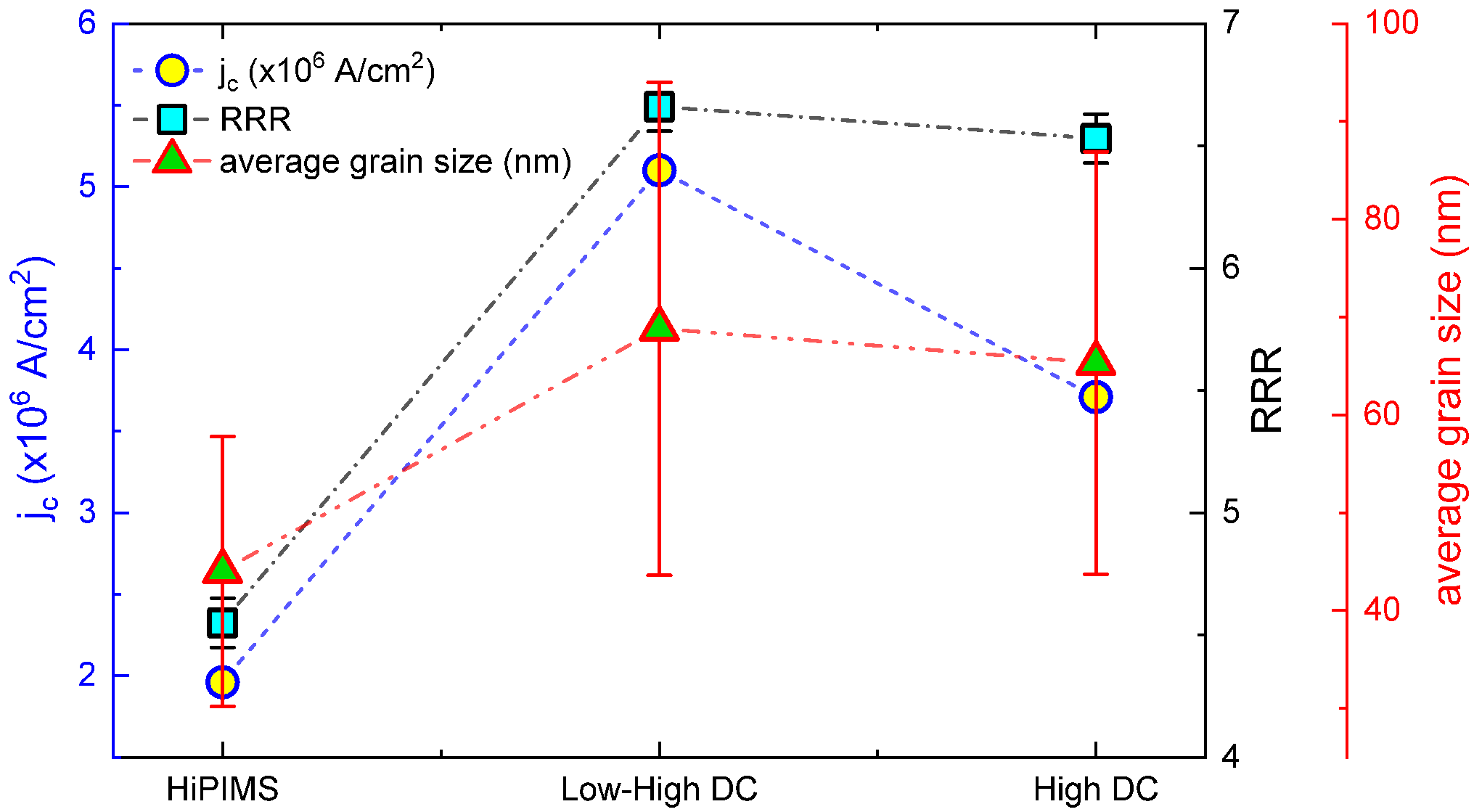} \caption{The estimated critical current density, residual resistivity ratio,
$RRR$, and the average grain size.}
\label{fig4:RRR_jc}
\end{figure}

\begin{figure}[tbh]
\includegraphics[width=1\linewidth]{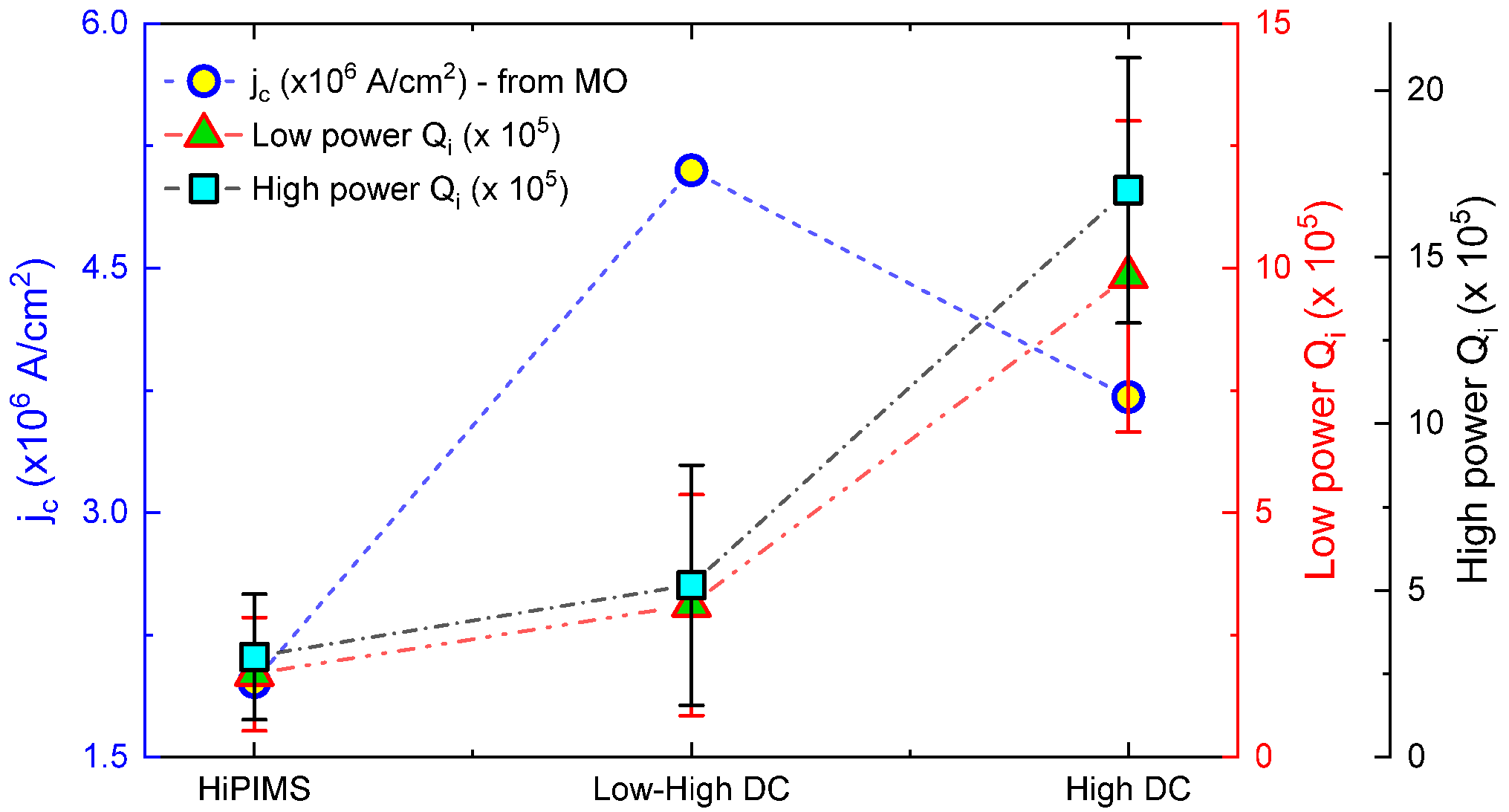} \caption{Side-by-side comparison of estimated critical current densities with
low and high power quality factor.}
\label{fig5:qi_jc}
\end{figure}

We can now correlate magneto-optical results with the physical parameters
of the studied films, such as the average grain size and $RRR=R(300\:\mathrm{K})$,
shown in Fig.\ref{fig4:RRR_jc}, as well as the internal quality factors,
$Q_{i}$ shown in Fig.\ref{fig5:qi_jc}. Sample A has the lowest values
of these parameters, sample B shows the largest, sample C shows intermediate,
but closer to sample B values. The higher value of $j_{c}$ in sample
B indicates higher pinning, seemingly contradicting the largest (but
still quite low) $RRR$. However, pinning strength depends on the
density of defects and the condensation energy. It is plausible that
the condensation energy term dominates. The normal state resistivity
is determined only by the defects. Significantly larger grain size
implies a low concentration of defects, consistent with our assumption. 

Next, we examine our results considering measured internal quality
factors. Figure \ref{fig5:qi_jc} shows that both low power and high
power quality factors progressively increase from sample A to sample
C. The lowest $j_{c}$, lower ability to screen the magnetic field,
and smaller grains are consistent with the lowest quality factor among
three types of films. On the other hand, sample B shows the highest
ability to screen a magnetic field and most robust superconductivity
as concluded above. Yet, its quality factor is not the highest. This
can be explained by referring to magneto-optical imaging, Fig.\ref{fig1:MO_4K}
and Fig.\ref{fig2:MO_6K}, where we observed that sample B is most
prone to dendritic thermo-magnetic instabilities, implying an impeded
heat conductance, which prompts us to examine the differences between
samples in the niobium/substrate interface. The TEM images show that
sample A has the thickest intermixed interface with a gradual transition
from niobium to silicon and increased metallicity \citep{oh2024}.
This results in a good thermal contact, which is consistent with the
observation that sample A did not show significant thermo-magnetic
instabilities in our experiments, Fig.\ref{fig1:MO_4K} and Fig.\ref{fig2:MO_6K}.
However, this improved thermal stability likely introduces a competing
drawback. A thicker, gradually intermixed interface reduces superconducting
order parameter, $\varDelta$. Along with smaller grains, which also
reduce it at their boundaries, this leads to a weaker superconducting
response, hence the lower quality factor. Judging by the wide range
of thermo-magnetic instabilities, sample B shows the weakest thermal
coupling, perhaps related to two different deposition velocities.
Finally, sample C hits the sweet spot - it shows excellent screening
capacity, comparable to sample B, but its thermal link to a substrate
is clearly better. Unlike other two films that show significant atomic
disturbance, while sample C is on the balanced side. This is likely
because this film has the thinnest spatially uniform silicide layer.
Naturally, sample C shows the best quantum performance.

\section*{Conclusions}

In conclusion, we used magneto-optical imaging to demonstrate that
the spatial inhomogeneity of the superconducting state and the capacity
to screen the applied magnetic field are important factors in determining
quantum coherence of Nb transmon qubits. Furthermore, the onset of
thermo-magnetic instabilities, observed as avalanches, shows the difference
in the thermal coupling to the substrate in the studied films. We
conclude that simultaneous optimization of the Nb/Si (and likely Nb/sapphire)
interfacial layer, the strength of the order parameter (higher
$T_c$, sharp superconducting transition) and flux pinning is needed to
achieve best results. The latter parameter, pinning, is particularly
delicate and must be studied and optimized at operating frequencies
of the device. Of course, other sources such as surface oxides, impurities,
two-level systems, chemical composition, and device geometry all contribute
to decoherence. Although one or more of these mechanisms may act as
a dominant bottleneck for a given film type and/or qubit design, it
is unlikely that a single source universally controls performance.
Systematic mitigation of each contributing loss channel is expected
to be required to achieve the best attainable device performance.

\begin{acknowledgments}

This work was supported by the U.S. Department of Energy, Office of Science, National Quantum Information Science Research Centers, Superconducting Quantum Materials and Systems Center (SQMS), under Contract No.89243024CSC000002. Ames Laboratory is supported by the U.S. Department of Energy (DOE), Office of Science, Basic Energy Sciences (BES), Materials Science \& Engineering Division (MSED) and is operated by Iowa State University for the U.S. DOE under contract DE-AC02-07CH11358. M.A.T. and K.R.J. were supported by DOE, BES, MSED at Ames National Laboratory. 

\end{acknowledgments}

%

\end{document}